\documentclass[showpacs,twocolumn,prl,aps,superscriptaddress]{revtex4}
\usepackage{amsmath}
\usepackage{amssymb}
\usepackage{graphicx}
\usepackage[normalem]{ulem}

\newcommand{\be}{\begin{equation}}
\newcommand{\ee}{\end{equation}}
\newcommand{\bea}{\begin{eqnarray}}
\newcommand{\eea}{\end{eqnarray}}

\newcommand{\Eq}[1]{Eq.\,(\ref{#1})}

\newcommand{\nl}{\nonumber \\}

\begin{document}
\title{Spin-Peierls-like transition in AFe$_2$As$_2$(A=Ba, Sr) }
\author{D. Hou}
\affiliation{Department of Physics, Renmin University of China,
Beijing 100872, P. R. China}\affiliation{School of Physics,
Shandong University, Jinan 250100, P. R. China}
\author{Q. M. Zhang}
\affiliation{Department of Physics, Renmin University of China,
Beijing 100872, P. R. China}
\author{Z. Y. Lu}
\affiliation{Department of Physics, Renmin University of China,
Beijing 100872, P. R. China}
\author{J. H. Wei}
\email{wjh@ruc.edu.cn} \affiliation{Department of Physics, Renmin
University of China, Beijing 100872, P. R. China}

\date{\today}

\begin{abstract}
From first-principles density functional theory calculations
combined with varying temperature Raman experiments, we show that
AFe$_2$As$_2$ (A=Ba, Sr), the parent compound of the FeAs based
superconductors of the new structural family, undergoes a
spin-Peierls-like phase transition at low temperature. The
coupling between the phonons and frustrated spins is proved to be
the main cause of the structural transition from the tetragonal to
orthorhombic phase. These results well explain the magnetic and
structural phase transitions in AFe$_2$As$_2$(A=Ba, Sr) recently
observed by neutron scattering.

\end{abstract}


\maketitle

{\it Introduction} The recent discovery of iron-based
high-transition temperature (high-$T_c$)
superconductors\cite{kamihara} has invoked great research
interests in similar materials with Fe-As layers. Currently, the
focuses are mainly on two kinds of structures: the first is iron
arsenide-oxides with P4/nmms space group \cite{X. H. Chen,G. F.
Chen,Z. A. Ren}, represented by the parent compound LaFeAsO;
 and the second is ternary iron-arsenide compound with body-centered
I4/mmm space group\cite{M. Rotter,C. Krellner, GF Chen} ,
represented by the parent compound BaFe$_2$As$_2$\cite{M. Rotter}.
Those compounds show similar structural and magnetic properties.
At room temperature, they all lay in non-magnetic, high-symmetric
state, with Fe-As layers separated by La-O layers or Ba layers
respectively. With the decrease of temperature, LaFeAsO undergoes
a slight structural transition from the tetragonal to orthorhombic
phase at $T_a=150$K, followed by the appearance of a magnetic
SDW(spin-density wave) state at 134K\cite{Clarina}, while
BaFe$_2$As$_2$ experiences a similar structural and magnetic
transition simultaneously at about $T_a=140$K \cite{bao}. At the
magnetic SDW state, they all form a collinear stripe-ordering
magnetic ground-state, with the nearest Fe atoms aligning
anti-ferromagnetically along one crystal axis in Fe-As plane,
while ferromagnetically parallel to the other axis\cite{J.
Dong,Clarina,bao,ma}. By doping with electron or hole carriers,
the structure transition and the SDW state are both suppressed,
and the superconductivity emerges at 52K\cite{Zhi-An Ren} and
38K\cite{rotter2} respectively. Thus there exists a competition
between magnetism and superconductivity in these compounds, but
the subtle details such as the pairing mechanism are far from
clear.

As that in high-$T_c$ cuprates, revealing the mechanism of
superconductivity of iron-based superconductors highly requires
understanding the electronic, structural and magnetic properties
of parent compounds first. The structural and magnetic phase
transitions seem to be common features in these Fe-As based
superconductors, thus revealing the sources of these transitions
and the possible connections between them may lead to better
understandings of the experimental observations. Some
first-principles density functional theory (DFT) calculations have
suggested that the frustrated superexchange interactions between
Fe ions induces the collinear stripe antiferromagnetic ground
state \cite{Yildirim,ma}, on the other hand the main cause of the
structural phase transition, as well as its correlation to the
magnetic one, is not well understood so far.

Let us start with the frustrated $J_1-J_2$ Heisenberg model to
describe the nearest neighbor and next-nearest neighbor
superexchange interactions among the Fe atoms bridged by As atoms,
which can be described as \cite{ma}
\be H=J_1\sum_{<ij>}\vec{S_i}\cdot\vec{S_j}+J_2\sum_{\ll
ij\gg}\vec{S_i}\cdot\vec{S_j},
\ee
where $<ij>$ and $\ll ij\gg$ denote the summation over the nearest
and the next-nearest neighbors respectively. The ground state of
the frustrated $J_1-J_2$ spin-half model on a square lattice at
zero temperature has been studied by several groups in the
literature and the main results are summarized as follows
\cite{Chandra88,Chandra90}: (1) In the absent of frustration
($J_2=0$), its ground state has long-range N\'{e}el order; (2)
With the increase of frustration ($J_2/J_1$), a phase transition
from N\'{e}el order to a spin-liquid phase occurs; (3) If further
increasing the frustration, a collinear order emerges at $J_2/J_1
\gtrsim 0.55$, with the nearest spins aligning
anti-ferromagnetically along one axis while ferromagnetically
parallel to the other.

According to the DFT calculation in Ref.~\onlinecite{ma}, $J_1$
almost equals to $J_2$ for Fe-As superconductors. In this sense,
the DFT and model calculations consistently explain the collinear
magnetic order of the parent compounds. However, the ground state
given by $J_1-J_2$ model is twofold degenerate with the $\pi/2$
rotational symmetry, which is not in agreement with the
orthorhombic structure observed in experiments. It indicates that
the pure spin model is not sufficient to account for the
structural phase transition. We hereby suggest to extend $J_1-J_2$
model by evolving the spin-phonon coupling. For examples, when
adiabatic phonons are considered, the model should be modified as,
\bea\label{JJP}H&=&\sum_{<ij>}[J_1(1-\alpha_1y_{ij})\vec{S_i}
\cdot\vec{S_j}+\frac{K_1}{2}y_{ij}^2]\nl%
&+&\sum_{\ll ij\gg}[J_2(1-\alpha_2y_{ij})\vec{S_i}\cdot\vec{S_j}
+\frac{K_2}{2}y_{ij}^2]\eea
where $y_{ij}=|\vec{u}_j-\vec{u}_i|$ with $\vec{u}_i$ denoting the
in-plane displacement of atom $i$, $\alpha$ is the spin-phonon
coupling constant. The ground state (at $T=0$) of above model was
calculated with a spin-wave approximation in recent literature and
a Peierls-like transition from a tetragonal to an orthorhombic
phase was found at large frustration ($J_2/J_1 \gtrsim
0.5$)\cite{Becca02}.

In this letter, motivated by the model analysis, we use the
first-principles density functional calculations combined with
varying temperature Raman experiments to study the ground state of
AFe$_2$As$_2$ (A=Ba, Sr). We prove that the spin-Peierls-like
phase transition is the very mechanism of the structural
transition at $T_a$ in the parent compounds of iron-based
superconductors.

\begin{figure}[htb]
\includegraphics[width=2.5cm]{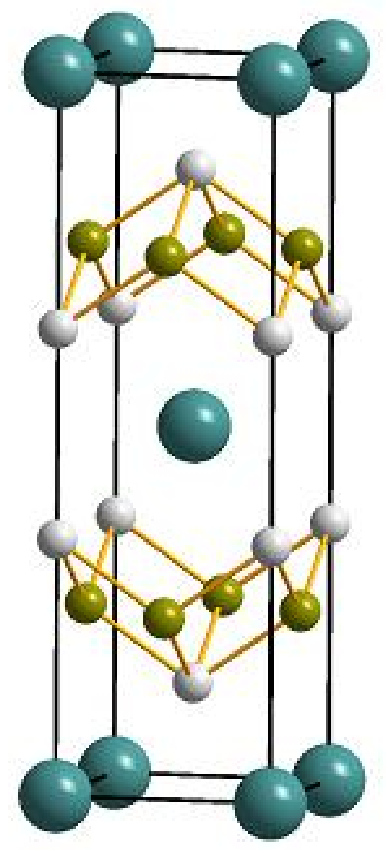}
\includegraphics[width=2.5cm]{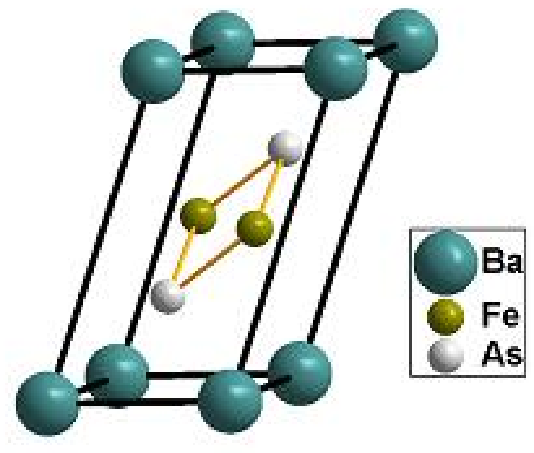}
\includegraphics[width=2.5cm]{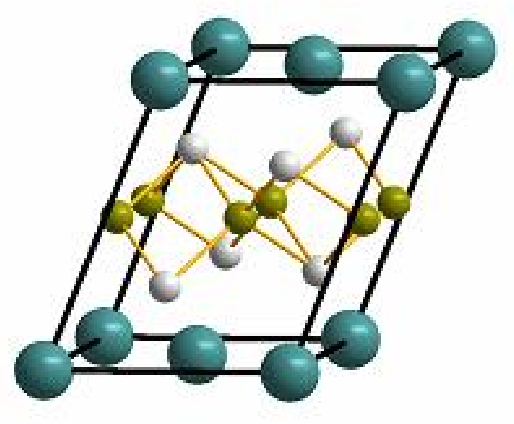}
\caption{(color online) Structure of BaFe$_2$As$_2$, from left to
right: conventional cell in I4/mmm space group, primitive
non-magnetic cell and primitive stripe-ordering cell} \label{fig1}
\end{figure}

{\it Method} The calculation was done using a plane-wave based
method\cite{QE} with local spin density approximation (LSDA) and
generalized gradient approximation (GGA) of Perdew-Burke-Ernzerh
(PBE)\cite{PBE} for the exchange-correlation potentials. The
density-functional perturbation theory (DFPT) was used to
calculate the $\Gamma$-point phonons. Firstly, the electronic
properties of AFe$_2$As$_2$(A=Ba,Sr) were calculated to determine
the electronic ground state, using experimental cell
parameters\cite{M. Rotter} and energy-minimized internal atomic
positions. Different magnetic configurations, namely the
nonmagnetic, square anti-ferromagnetic and collinear
anti-ferromagnetic were considered\cite{ma}. The ground states of
both BaFe$_2$As$_2$ and SrFe$_2$As$_2$ are found to be collinear
anti-ferromagnetic with a slight structure transition, which is in
agreement with previously reported results. For the phonon
calculations, we focus on the nonmagnetic high-temperature state
and the collinear anti-ferromagnetic ground state at low
temperature. Here we adopt the triclinic primitive unit cells as
shown in Fig.~\ref{fig1}, in which the structure transition alters
the angle between x and y crystal axes from 90$^\circ$ to
89.6$^\circ$.

The Raman measurements were performed with a triple-grating
monochromator (Jobin Yvon T64000), which works with a microscopic
Raman configuration. A 50$\times$ objective microscopic lens with a
working distance of 10.6 mm, is used to focus the incident light on
sample and collect the scattered light from sample. The detector is
a back-illuminated CCD cooled by liquid nitrogen. An solid-state
laser (Laser Quantum Torus 532) with high-stability and very narrow
width of laser line, is used with an excitation wavelength of 532
nm. The laser beam of 3 mW was focused into a spot of less than 10
microns in diameter on sample surface.

{\it Result} From symmetry analysis, the Raman phonon of
nonmagnetic AFe$_2$As$_2$ with I4/mmm space group consists of four
modes: the A$_{1g}$ and E$_g$ for As, the B$_{1g}$ and E$_g$ for
Fe. The vibrating directions in A$_{1g}$ and B$_{1g}$ modes are
perpendicular to the Fe-As layer, while that of E$_g$ parallel to
the Fe-As layer. After cooling down to the transition temperature
$T_a$, the crystal structure changes into orthorhombic F/mmm space
group, and the magnetism appears almost at the same temperature.
In this space group, the E$_g$ modes split into B$_{2g}$ and
B$_{3g}$ modes, and A$_{1g}$ mode for As changes into B$_g$ mode.
The calculated Raman modes, as listed in Table 1, are in good
agreement with symmetry analysis. All of the modes have been found
and assigned from atomic displacements.

The Raman modes of SrFe$_2$As$_2$ have been measured in
Ref.\onlinecite{A.P. Litvinchuk} but only for nonmagnetic state
reported. By comparing our calculated phonon frequencies with the
experimental values, we find a systematic frequency shift ($\sim
20$cm$^{-1}$) of the calculated values towards higher value region
(see Table 1) due to the temperature effect. That is to say, the
DFPT theory deals with the ground-state problems, which in this
case corresponds to a non-magnetic state at $T=0K$, while the
Raman spectra was measured at room temperature, as a consequence,
the corresponding experimental values are lower.

Since the strong spin-phonon interaction is the basis of the
suggested mechanism of spin-Peierls-like phase transition, we must
verify this point first, by calculating the change of the Raman
modes in nonmagnetic state without the magnetic transition but
with the structural transition. In that case, the calculated Raman
modes and frequencies almost remain unchanged for both
BaFe$_2$As$_2$ and SrFe$_2$As$_2$, with frequency shifts (or
splits) less than 3 cm$^{-1}$. This suggests that the slight
structure transition has limited effects on the electronic
properties and Raman phonons. Our tentative calculations also
gives us another valuable information: the E$_g$ mode of Fe atom
experiences a frequency splitting as a result of the structural
transition, which consists with the fact that the structure
distortion mainly appears in Fe-As layer and changes the nearest
neighbor Fe-Fe distances slightly, so that the nearest neighbor
Fe-Fe distances are no longer equal in the orthorhombic crystal
axes directions.

\begin{widetext}
\begin{center}
Table 1: Raman phonon frequencies of AFe$_2$As$_2$(A=Ba,Sr) in the
non-magnetic state (marked as N) and in the collinear
anti-ferromagnetic state (marked as C). The Raman modes are in
I4/mmm space group corresponding to the non-magnetic state, while
those in the brackets are in P/mmm space group for the collinear
anti-ferromagnetic state.The measurement temperatures are 260K and
87K for SrFe$_2$As$_2$, and 290K and 100K for BaFe$_2$As$_2$.
The atomic displacements are for the collinear anti-ferromagnetic state.\\
\begin{tabular}{|c|c|c|c|c|c|c|c|c|}\hline\hline
Atom & Raman Mode& SrFe$_2$As$_2$-N & SrFe$_2$As$_2$-C &
BaFe$_2$As$_2$-N & BaFe$_2$As$_2$-C &
 Displacement\\
     & (cm$^{-1}$) & Cal. (Exp.) & Cal. (Exp.) & Cal. (Exp.) &  Cal. (Exp.)
     &
     of Atoms, C
     \\\hline
As & E$_g$(B$_{2g}$) &138.9&96.1 & 140.2  & 86.8  & As(x), Fe(y)\\
As & E$_g$(B$_{3g}$) &138.9 &126.3 &140.2 & 125.6 & As(y), Fe(x)\\
As & A$_{1g}$(B$_g$) &207.6 &183.4 & 205.7&  177.8 & As(z)\\
Fe & B$_{1g}$(B$_{1g}$)&219.5(206)&219.3(213)& 224.0(206)&  214.2(212)  & Fe(z)\\
Fe & E$_g$(B$_{2g}$) &301.2 &262.4&  293.6 & 252.0 &Fe(y), As(x)\\
Fe & E$_g$(B$_{3g}$) &301.2 &289.0&  293.6 &  281.1 &Fe(x),
As(y)\\\hline \hline

\end{tabular}
\end{center}
\end{widetext}

\begin{figure}[tb]
\includegraphics[width=6cm]{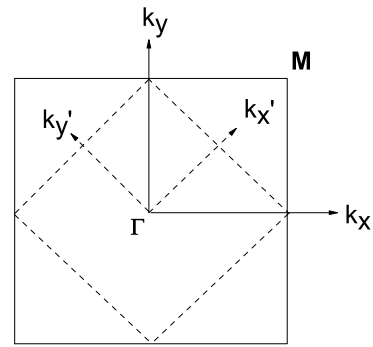}
\caption{Reciprocal space of BaFe$_2$As$_2$ in non-magnetic(solid
lines) and stripe-ordering(dashed lines) state.} \label{fig2}
\end{figure}

We are now on the position to elucidate the changes of the Raman
modes due to spin-phonon interactions at the magnetic ground
state. Our calculations on the electronic structure verify that
the magnetic ground state of AFe$_2$As$_2$(A=Ba,Sr) is a collinear
stripe-ordering anti-ferromagnetic one. The Fe spins align
parallel along the shorter axis in Fe-As layer and anti-parallel
along the longer one, as observed by neutron diffraction
measurement\cite{bao}, which is similar to that of
LaFeAsO\cite{Clarina}. To correctly illustrate this magnetic
configuration, the nonmagnetic primitive unit cell should be
doubled in a-b plane to include two formula cells. Here we adopt a
$\sqrt{2}$$\times$$\sqrt{2}$$\times$1 cell, with x(y) axis rotates
45$^\circ$ and points to the nearest Fe-Fe directions. By doubling
the crystal lattice, the reciprocal lattice rotates and shrinks to
a half, and the original reciprocal points fold down to the new
points. The folding is illustrated in Fig.~\ref{fig2} (the
structure distortion from tetragonal to orthorhombic not shown in
the figure). It is clear that original M($\pi$, $\pi$, 0) points
fold down to the new $\Gamma$(0, 0, 0) point. Thus the obtained
$\Gamma$ point phonon modes in SDW state doubly contain the
information from both original $\Gamma$ and M points. Only the
modes corresponding to $\Gamma$ point are picked up according to
the atomic displacements, which are shown in Table 1 (marked as
C). For the purpose of comparison, we also show the measured Raman
phonons at different temperatures before and after the magnetic
transitions in the table. The calculated phonon frequencies are
found very close to the experimental values, which further proves
the reliance of our phonon calculations.

As shown in Table 1, accompanied with the onset of magnetism,
almost all of the Raman phonon frequencies softened to lower
values, specifically some of them (As-B$_{2g}$ and Fe-B$_{2g}$)
show a giant phonon softening (GPS). The GPS of Raman phonon modes
is the main result of our DFT calculations, which directly support
the mechanism of the spin-Peierls-like phase transition at $T_a$
in AFe$_2$As$_2$(A=Ba, Sr). Let us elaborate this point in more
details as follows:

Firstly, by comparing the changes of B$_{3g}$ and B$_{2g}$
frequencies of Fe atoms (seen in Table 1), one can see that the
former only reduces 12cm$^{-1}$ for both BaFe$_2$As$_2$ and
SrFe$_2$As$_2$ while the latter reduces 41cm$^{-1}$ for
BaFe$_2$As$_2$ and 39cm$^{-1}$ for SrFe$_2$As$_2$ (GPS). The
difference between these two modes is the involved Fe atoms
vibrating in different direction - along x and y crystal axis
respectively. In SDW state, the spins on Fe atoms align
antiparallel along x (longer) direction, while parallel along y
(shorter) direction. In B$_{2g}$ mode, the vibration direction of
Fe atoms accords with the parallel spin direction, suggesting an
instability of parallel alignment of Fe spins.

Secondly, if assuming $y_{<ij>}^{x,y}=\delta^{x,y}$, $y_{\ll
ij\gg}^{x,y}=\delta^x+\delta^y$, $\alpha_1=\alpha_2=\alpha$ and
$K_1=K_2=K$ in the extend $J_1-J_2$ model [see \Eq{JJP}], one can
obtain the collinear state (ground state) with
$\delta^x=\alpha(J_2-J_1)/4K$ and $\delta^y=\alpha(J_2+J_1)/4K$ at
sufficient large frustration $J_2/J_1>0.5$ \cite{Becca02}. In
consideration of $J_2\sim J_1$ in AFe$_2$As$_2$, one can reach
$\delta^y\gg\delta^x$, which consistently explains why the GPS of
Raman modes of Fe atoms mainly happens along y crystal axis.

Thirdly, the spin-Peierls-like phase transition proved by the DFT
and model analysis here closely relates to the "exchange
striction" effect that also predicts symmetry breaking distortions
from the magnetoelastic interaction\cite{goodenough}. That effect
is believed to play an important role for the origin of the
multiferroics, and furthermore, we prove here it may be also
related to the high-T$_c$ superconductivity.

{\it Summary} In summary, the consistence of the first-principles
density functional calculations and model analysis strongly prove
that the spin-Peierls-like phase transition is the very mechanism
of the structural transition at $T_a$ in parent compounds of
iron-based superconductors. We thus conclude that the dominate
interactions in those parent compounds are frustrated spin-spin
interaction and spin-phonon interaction. Electron or hole doping
suppresses the ground state of parent compounds (orthorhombic
structural and collinear magnetic phase) and induces the
superconductor state. In order to understand the pairing mechanism
for ion-based new superconductors, one should treat the spin-spin
and spin-phonon interactions on an equal footing.

Supports from the National Natural Science Foundation of China
(Grants No.~10604037) and the National Basic Research Program of
China (Grants No.~2007CB925001) are gratefully acknowledged.

\end{document}